# バーチャルツイン設計論開発に向けた
# 場所再現の魅力に関するインタビュー調査

青柳 西蔵[*1]

## Interview Survey on Attractivenesses of Place Re-creation Toward Developing a Virtual Twin Design Theory

Saizo Aoyagi[*1]

**Abstract** – It is often seen that real-world locations are re-created using models, metaverse technology, or computer graphics. Although the surface-level purposes of these re-creations vary, the author hypothesizes that there exists an underlying common attractiveness that remains unclear. This research aims to clarify the attractivenesses and its structures of place re-creations through an interview study with qualitative analysis. The interviews used examples of physical re-creations, such as the model in Komazawa University's Zen Culture History Museum and some dioramas of Tokyo, as well as computer-generated re-creations of Shibuya using platforms like Minecraft and Project Plateau's 3D city model. Using insights gained from this investigation, this study seeks to establish a theoretical framework for designing virtual twins.

## 1. はじめに

近年バーチャルツインの利用が広がっている。バーチャルツインとは、ある現実の場所をバーチャル環境として再現したものである。ユーザはその中を自由に歩き回ったり、物体とインタラクションしたりできる。これはデジタルツインの一種であり、3DCG (三次元コンピュータグラフィックス) モデルを用いたバーチャルリアリティ（VR）の体験をユーザに提供するものである [1]。バーチャルツインの規模としては、手に持てる物体から大都市まで様々であるが、本研究では、人が歩き回ることのできる街くらいの大きさのバーチャルツインに焦点を当てる。こうした例は、歴史教育コンテンツ[2]、ゲーム内のステージ[3]、メタバース[4]等、多数みられる。また著者は、将来的に現実の場所を完全に代替できるバーチャルツインを実現することで様々な社会的問題を解決するという構想を持っている[5]。

しかし、バーチャルツインを成功に導くための設計論は未だ確立していない。まず、商業的な事例が多いため明示的な設計論が示されていないことが多い。また示されている場合でも、3Dスキャンデータを用いる等により単に現実と客観的に忠実にすること目標にする例が多い[6,7]。これに対し著者は、人が持つ認知マップや都市イメージ等の場所の主観的な表象[9,10]とバーチャルツインの体験を一致させる設計論が必要であると考えている。

具体的に、場所のどのような点をどのように再現すべきなのだろうか？これに答えることが設計論の確立につながる。そのために、本研究では場所再現の魅力に着目する。場所再現はバーチャルツインのみならず現実世界における再現も含めた概念である。例えば、伊勢神宮の式年造替や江戸深川の三十三間堂といった写し建築[11]、大阪城等の復元城郭、模型やジオラマ展示[12]、世界の有名建築を再現したテーマパーク[13]、等がある。場所再現はバーチャル技術発明以前から、古来作られ続けている。

これらの場所再現には、教育や娯楽等、それぞれに表層的な目的がある。しかし著者は、場所再現には共通した特有の魅力とそれを実現する要因や機構があると推測している。これらを把握することで、その機構を利用してその魅力を強化するという方向で、バーチャルツインの設計論を開発する。これが本研究を含む本研究プロジェクト全体のアプローチである。

これまでの研究で文献調査を通して、写し[14]や見立て[15]といった日本文化に見られるポジティブに評価される模倣、遊び研究[16]、フロイトによる不気味さの概念[17]等との場所再現の関連を指摘した[12]。これに基づき、場所再現の魅力は(1)違うものの間に共通性を見出す見立て、(2)現実のツラさを排除し面白いところだけを取り入れた遊び、(3) 親しみと異質さの同居が生み出す不気味を楽しむホラーにある、という3つの仮説を見出した[18]。

---

*1: 駒澤大学　グローバル・メディア・スタディーズ学部
*1: Faculty of Global Media Studies, Komazawa University

しかし、これらの仮説は実際の場所再現のユーザの声に基づくものではなく、そこに乖離がある可能性もある。著者は、これらの仮説検証に進む前に、ボトムアップにユーザの体験から改めて場所再現の魅力についての仮説を生成することで、研究を補強したいと考えた。

本研究では、実際に場所再現に魅力を感じている人々から、どのような点が魅力なのかについてインタビューによって調査する。その結果を基に場所再現の魅力の概念モデルを作成することが本研究の目的である。本研究の貢献は、ユーザの声に基づき、特定の場所の特定の形式の再現についてではなく、場所再現一般の魅力に関する概念モデル(仮説)を生成することである。

## 2. 関連研究

場所再現を扱い、その体験価値を調べた質的研究の事例はいくつか見られる。アメリカ南北戦争の再現の研究は、戦場の一つであったゲティスバーグでの再現のインタビュー調査から、再現の参加者によって集団的に真正性が達成されていると論じた[19]。ポルトガルにおける戦場観光の調査研究は、再演や歴史的レクリエーションのような遺産をテーマとした観光イベントが地域のアイデンティティと記憶を強化すると分析した[20]。都市計画分野の中国の研究は、仮想空間と物理空間の両方を用いて過去を象徴的に再現することで場所のアイデンティティを形成する新ノスタルジック空間という概念を提唱した[21]。アニメ聖地巡礼[22]はアニメ上での現実の場所の再現があり、それの再現であるかのように再現元を訪れるというやや複雑な場所再現の例である。大津市と宇治市の観光の研究では、ファンが自ら情報を検索し、聖地で作品と同じ角度で写真を撮るといった行動をとり、地域の人達やファン同士の交流を通してみずから体験価値を共創していくと論じた[23]。これらは、いずれも現実世界の対象の場所自体に再現という要素を持ち込む場合の体験価値の例である。

他方、場所を元の場所とは別の形で再現する形態としてジオラマ模型の体験価値を扱った研究も多い。博物館への来館者のジオラマの影響についての研究では、ジオラマは動物園より自然に近く、それを観察した来館者は生物学的・生態学的により正確な自然の知識を得て、自然保護の意識が高まるとした[24]。一方で、ライオン生息地のサバンナの風景のジオラマを題材に来館者に自ら写真撮影させる手法とインタビューを組み合わせて鑑賞体験を調べた研究では、来館者は展示の意図を理解しつつ、展示を批判的に見て独自の物語を構築すると示した[25]。また、ジオラマにより精巧で現実感のある情景を目の当たりにすることで「まるでその場にいるような」没入体験が促され、来館者は過去に実体験した自然の記憶や郷愁を呼び覚まされることを示したVR研究のような観点の例もある[26]。

上記の研究は事例研究がほとんどである。例外的に、デジタル技術とも融合したジオラマには観察力や問いかけを促す効果があるとするレビュー研究があるが、これは博物館展示の一形式という観点に終始している[27]。これらに対する本研究の特徴は、現実世界におけるものもバーチャルなものも含めた、場所再現一般に対する体験価値を対象とする新しい試みであることである。

## 3. 調査

### 3.1 調査の目的と方法

場所再現の魅力についての意見を広く調べる、半構造化インタビュー調査を実施した。本調査は駒澤大学「人を対象とする研究」に関する倫理委員会の承認を受けた(審査番号: 25-6)。

参加者は「模型、マイクラの再現、再現テーマパーク、メタバース的なバーチャル○○等」、「実在の場所を再現したものが好きな人」、「ちょっと好き、くらいでもいい」という文言で募集した。参加者は駒澤大学の学生7名であった。彼らは、女性と男性1名ずつのグループ1、男性2名のグループ2、女性2名と男性1名のグループ3に分かれてそれぞれ調査に参加した。

本調査はグループごとに、図1に示す流れで行った。(1)参加者は、最初に調査内容について説明を受け、参加に同意した。(2)これまでに場所再現を見たり訪れたりした経験の内容を説明した。(3)次に、実世界の場所再現の2件を閲覧した。例として用いたのは、駒澤大学の禅文化歴史博物館[*3]にある大学周辺の模型を撮影した写真(図2(a))、動画、3Dスキャン(図2(b))と、第2回東京ジオラマ展のWebサイト[12]に掲載された新宿駅や三軒茶屋駅の模型の写真である(図3)。ただし、グループ3のみ実際に博物館を訪問し模型を閲覧した。(4)また、バーチャルな場所再現の例2件を閲覧し、またグループで一組のXboxゲームコントローラ、マウス、キーボードを共有して自由に再現内を移動した。最初に用いた例は

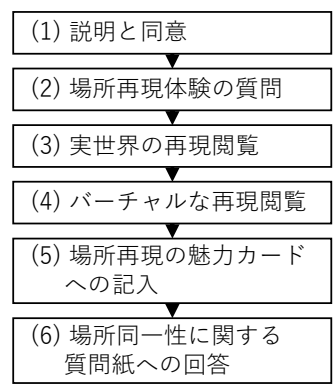

図1 インタビュー調査の手順
Fig.1 Steps of the Interview Survey.

---

*3: https://www.komazawa-u.ac.jp/facilities/museum/
*4: https://www.minecraft.net

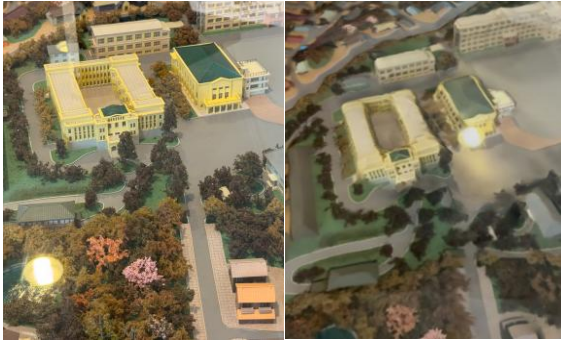

(a) 写真　　　　　(b) 3D スキャン

図 2: 駒澤大学の禅文化歴史博物館の模型
Fig.2 The Model in Komazawa University's Zen Culture History Museum.

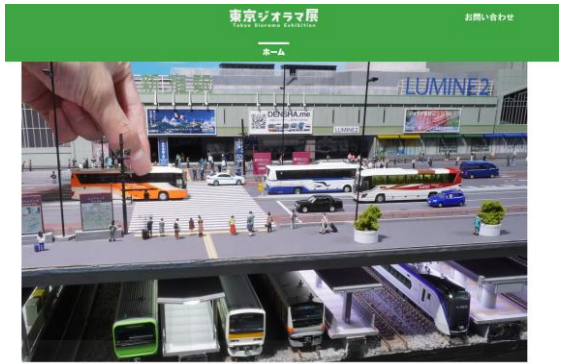

図 3: 東京ジオラマ展の Web サイト[13]
Fig.3 The Website of Tokyo Diorama Exhibition[13].

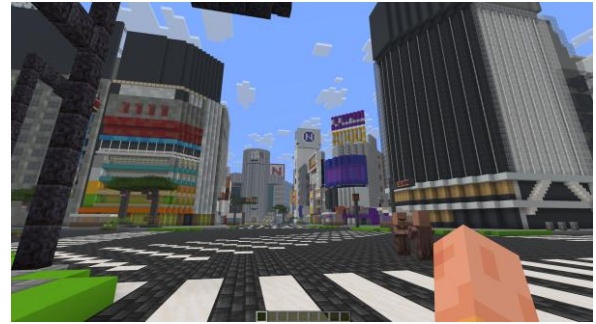

図 4: Noshychan による渋谷の Minecraft 再現[15]
Fig.4 Minecraft Re-creation of Shibuya by Noshychan[15].

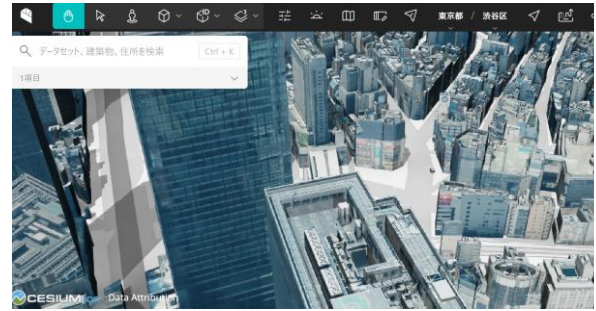

図 5: PLATEAU VIEW で見る渋谷
Fig.4 Shibuya Shown in PLATEAU VIEW.

Microsoft 社の箱庭ゲーム Minecraft*4 を用いて作られた Noshychan による渋谷の再現[28]である(図 4)。これを Minecraft のゲーム上に読み込んで閲覧した。次の例は、国土交通省による Project Plateau[29]の 3D 都市モデルの渋谷区のデータである(図 5)。これを Web ブラウザのビューア PLATEAU VIEW 4.0 で閲覧した。(5) 次に、「場所再現にはどのような魅力があるか」を考えて、思いついたものを正方形型のふせんに 1 つずつ書き、台紙となる白い画用紙に貼り付けた。

(6)最後に場所同一性に関する質問紙に回答した。これは Google Form を用いて作成されたもので、スマートフォンで回答した。これに含まれる項目は、これまでの研究で用いてきた、場所再現がどの程度元の場所と同じと感じるかを表す概念、「場所同一性」を測定する質問紙項目[5]と、場所同一性に関連する諸概念[11]を測定する質問紙項目である。場所アイデンティフィケーション [30]、場所の印象 [31]、場所の真正性 [32,33]、空間的存在感[34]、親近性[35]については既存尺度を日本語訳して用いたが、答えるのが困難と感じるほど項目数が多くなってしまったので 1 概念あたり 3 項目以下になるように一部を削減した。またノスタルジア[36]、不気味さ[37]、写し[14]、見立て[15]、複数の実体化や型の出来事の同一性[38]については既存研究を参考に著者が項目を作成した。参加者は、この調査中に取り上げた中で最も印象に残った場所再現を一つ選び、それに対する評価について回答をした。本研究は仮説生成を目的としており、これらの測定によって仮説検証をするものではなく、あくまで得られた結果の解釈の補助として用いる。

これらの手順全体を通し、調査者は適宜参加者の発話を引き出す質問をし、また参加者はその直接的な回答に限らず他の参加者や調査者と自由に話した。そして、同意を得た以降の全ての発話及び調査風景を録音・録画した。なお、調査時間は 1 グループあたり 90 分であった。図 1 の各手順にかけた時間はその時のインタビューの流れに依存して各グループで異なった。

### 3.1　結果

調査の結果、インタビューの発話データが約 240 分、場所の魅力をカードを書いたもの(魅力カード)が 33 枚、場所同一性に関する質問紙への回答 7 名分が得られた。本稿執筆時点ではまだ分析が完了していないが途中経過について説明する。

表 1 魅力カードをまとめた KA カードの一部例
Table 1 KA Card Examples Extracted from Attractiveness Cards.

| 出来事 | 心の声 | 価値 |
|---|---|---|
| (行為)その場所にいかなくても見れることで(結果)事前情報を得られる。 | 行かなくても気になるあの場所が見られる！ | ある場所に実際に行かずに家から見ることのできる価値 |
| (原因)単純に作りが細かくて(行為)見ていて(結果)おもしろい引き込まれる | 詳細な制作物はお面白い！ | 作りの細かい制作物を見ることのできる価値 |
| (原因)ジオラマ等を用いると、(行為)おなじ「再現」でも模型・メタバース等、媒体による違いを見比べられる。(結果)それは楽しい。 | おなじ「再現」でも模型・メタバース等、媒体による違いを見比べるのも楽しい。 | 異なる媒体による再現を見ることができる価値 |
| (原因)バーチャル世界での再現だと、(行為)時間を自由に設定できて、色々な景色が見られる | バーチャル世界での再現だと、時間を自由に設定できて、色々な景色が見られる | 場所の時間を自由に設定できる価値 |

　魅力カードの内容を、ユーザエクスペリエンスデザイン分野で製品やサービスに関するユーザの体験価値を抽出する質的研究手法である KA 法[39,40]を用いて分析した。この方法では、得られたデータを項目の内容からなる「KA カード」と呼ばれる形式でまとめる。まずユーザの行為を主な内容とした「出来事」を抽出し、そこからユーザの心境を想像して「ユーザの心の声」を端的に表現する。最後に出来事と心の声を手がかりにその理由や意味を解釈して「価値」を書く。結果の分析より得られた KA カードの一部を表 1 に示す。

　KA 法では、得られた多数の KA カードを類似するもの同士をまとめて「中分類の価値」を作り、それらの間の因果関係や経時的な関係を見つけて構造化し「価値マップ」という図としてまとめる。現時点での価値マップを図 6 に示す。

### 3.2　考察

　分析の結果、物理的に移動しなくてもその場所を疑似的に訪れることができる「閲覧の楽さ」、俯瞰視点や過去のある時点など時空間的に「現実の場所を異なる視点で見られる」、元の場所との比較によって成り立つ「再現と現実の差異が面白い」、現実の場所に含まれない追加情報などが持つ価値である「再現特有の良さ」の 4 つの価値を抽出した。元の場所が基から持つ価値を補強する価値と、再現をすることによって新たにできる価値の両方があるのが興味深い。

　なお、KA 法で分析し価値としてまとめた内容と元のカードに記載の内容の違いがあまりなかった。結果として

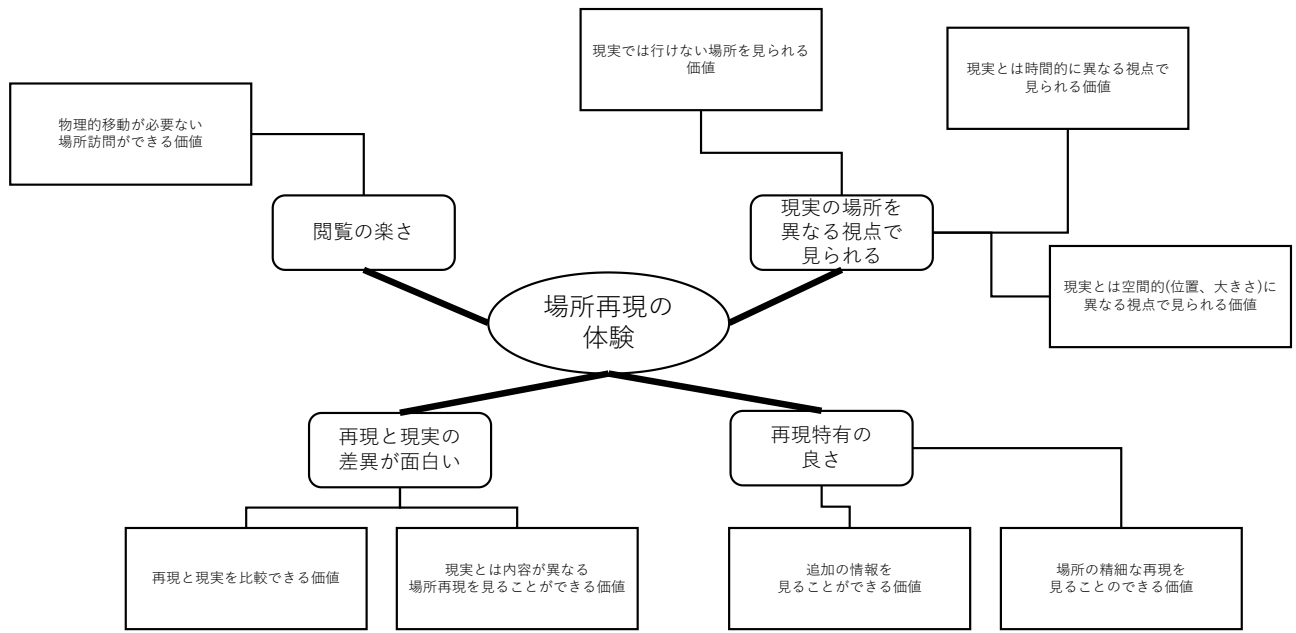

図 6 価値マップ
Fig.6 The Value Map

KA法を実施することの意義が小さいと考えられた。これは、場所の魅力カードは、調査の参加者自身が自分の心の声をまとめた内容であるためと考えられる。KA法は、インタビュー発話内容のような、構造化やまとめがされていない自然かつ自由なテキストデータに対して効果を発揮する分析手法であると推察される。ただし、著者の分析への不慣れさが結果に影響したことも否めない。今後も分析を深めていく必要がある。

## 4. おわりに

本研究では、実際に場所再現に魅力を感じている人々から、どのような点が魅力なのかについてインタビューによって調査した。その結果の分析の途中経過の報告として魅力カードのKA法によるまとめを示した。今後は、発話データを文字起こしし、質問紙の結果を参考にしながら魅力カードと共に解釈して、場所再現の魅力がどこにあるか概念モデルとしてまとめる。また、これを基にバーチャルツインのデザイン要件を抽出する予定である。

より長期的な観点では、現実世界の場所において他の場所を再現する例等、より広い範囲の場所再現を研究に取り入れたい。